\documentclass[11pt,a4paper]{article}

\usepackage[utf8]{inputenc}
\usepackage[T1]{fontenc}
\usepackage{lmodern}
\usepackage{microtype}
\usepackage[margin=2.5cm]{geometry}
\usepackage{setspace}
\usepackage{parskip}
\usepackage{graphicx}
\usepackage{booktabs}
\usepackage{enumitem}
\usepackage{hyperref}
\hypersetup{colorlinks=true,linkcolor=blue,citecolor=blue,urlcolor=blue}
\usepackage{titling}
\usepackage{fancyhdr}
\usepackage{caption}
\usepackage{amsmath,amssymb}
\usepackage{setspace}
\usepackage{multicol}
\usepackage[numbers,sort&compress]{natbib}
\usepackage{wrapfig}   % in preamble

\begin{document}

%%%%%%%%%%%%%%%%%%%%%%%%%%%%%%%%%%%%%%%%%%%%%%%%%%%%
%               COVER PAGE
%%%%%%%%%%%%%%%%%%%%%%%%%%%%%%%%%%%%%%%%%%%%%%%%%%%%

\begin{titlepage}
    \centering
    \vspace*{2cm}

    {\Huge\bfseries Expanding Horizons \\[5pt] \Large Transforming Astronomy in the 2040s \par}
    \vspace{1.5cm}
    {\LARGE \textbf{Towards a Comprehensive Understanding of Planetary Systems through Population-Level, Large-Scale Surveys}\\[0.9cm]}

{\large
Francisco J. Pozuelos$^{1,\star}$, Pedro J. Amado$^{1}$, Jesús Aceituno$^{2}$, Marina Centenera-Merino$^{1}$, Stefan Cikota$^{2}$, Javier Flores$^{2}$, Julius Göhring$^{3}$, Sergio León-Saval$^{4}$, Kalaga Madhav$^{3}$, Giuseppe Morello$^{1}$, Abani Nayak$^{3}$, Jose L. Ortiz$^{1}$, David Pérez-Medialdea$^{1}$, María Isabel Ruiz-López$^{1}$, Miguel A. Sánchez-Carrasco$^{1}$, Alejandro Sánchez-López$^{1}$\\[0.5cm]
\vspace{0.1cm}

    $^{1}$\textit{Instituto de Astrof\'isica de Andaluc\'ia, CSIC, Glorieta de la Astronom\'ia s/n, 18008 Granada, Spain}\\
    $^{2}$\textit{ Centro Astronómico Hispano en Andalucía, Observatorio de Calar Alto, Sierra de los Filabres, 04550, Gérgal, Almeria, Spain}\\
    $^{3}$\textit{Astrophotonics (innoFSPEC), Leibniz-Institut für Astrophysik Potsdam (AIP), Germany; Institut für Physik und Astronomie, Universität Potsdam, Germany}\\  
    $^{4}$\textit{Sydney Institute for Astronomy, The University of Sydney, School of Physics A28, Sydney, NSW 2006, Australia}\\
    $^{\star}$\textit{pozuelos@iaa.csic.es}
    }

    \vfill

    {\large
    ESO Call for White Papers on Future Science Questions\\[0.2cm]
    Submission date: 15 December 2025
    }

    \vspace*{2cm}
\end{titlepage}

%%%%%%%%%%%%%%%%%%%%%%%%%%%%%%%%%%%%%%%%%%%%%%%%%%%%
%        MAIN TEXT (MAX 3 PAGES, 11PT)
%%%%%%%%%%%%%%%%%%%%%%%%%%%%%%%%%%%%%%%%%%%%%%%%%%%%

Over the past three decades, exoplanet research has progressed from the first detections to a rapidly expanding census that now exceeds 6,000 confirmed planets spanning a wide range of masses, sizes, and orbital configurations. These discoveries have fundamentally transformed our view of planetary systems, revealing populations that occupy regions of parameter space with no direct Solar System analogues \citep[see, e.g.,][]{howard2013,winn2015}. However, the physical interpretation of this census remains severely limited. 
Robust constraints on planetary interiors and atmospheres require precise, jointly determined bulk properties, typically better than $\leq$5\% in radius and $\leq$15\% in mass, to break degeneracies among composition, structure, and evolutionary state \citep[see e.g.,][]{dorn2015,plot2024,acuña2024}. Despite the large number of known exoplanets, only a small fraction currently meet these requirements: about 10\% of giant planets and less than 1\% of planets smaller than $\sim$4 Earth radii (NASA Exoplanet Archive on December 2025). Consequently, the vast majority of known exoplanets lack the physical characterisation needed to constrain their internal structures and atmospheric properties meaningfully.

This disconnect highlights a fundamental limitation of the current exoplanet landscape: although detections are abundant, most planets remain physically poorly understood, preventing the construction of robust population-level trends and limiting our ability to identify the physical processes that govern planet formation, evolution, and habitability across planetary systems.

\subsection*{Limits of detection-driven exoplanet surveys}

Space-based transit surveys have revolutionised exoplanet science by mapping large regions of parameter space in planetary radius and orbital period. Missions such as Kepler/K2 \citep{kepler} and TESS \cite{ricker2015} have established the basic architecture of planetary populations, demonstrating that planets are ubiquitous across a wide range of orbital configurations. In the coming years, this census will expand dramatically with new facilities, including ESA’s PLATO mission \citep{plato2025}, the Earth 2.0 concept \cite{earth2022}, and the Nancy Grace Roman Space Telescope, which together will deliver an unprecedented inventory of planets across stellar types, orbital periods, and planetary sizes.

PLATO will provide precise planetary radii and orbital properties for thousands of planets around bright FGK stars, supported by accurate stellar characterisation via asteroseismology. Earth 2.0 will extend demographic studies through combined transit and microlensing surveys, probing complementary regimes such as wider-orbit and free-floating planets, while Roman will further expand this landscape, particularly around low-mass and late-type stars. Collectively, these missions will densely populate the radius–period plane, enabling detailed studies of occurrence rates across a broad range of planetary and stellar parameters.

Despite this progress, a fundamental limitation will persist. For the vast majority of detected planets, these surveys will provide little more than planetary radii and orbital periods, while direct constraints on masses, interior structures, and atmospheres will remain available only for a small subset of favourable targets, leaving many planets physically ambiguous.

This limitation does not stem from insufficient instrumental precision, but from the prohibitive observing time required for large-scale follow-up. Although state-of-the-art high-resolution spectrographs on 5–8 m class telescopes can achieve the precision needed to measure the masses of even temperate Earth-sized planets orbiting solar-like stars, achieving mass accuracies better than $\sim$15\% typically requires 800–1200 radial velocity measurements per target. Such investments are incompatible with the multi-science nature and competitive time allocation of current and forthcoming facilities. Consequently, without systematic mass measurements and interior constraints at scale, large regions of the observed radius–period parameter space will remain degenerate, limiting our ability to distinguish their actual nature.

\subsection*{Atmospheric characterisation beyond individual targets}

Exoplanet atmospheric studies have recently entered a new era. The James Webb Space Telescope (JWST) has delivered unprecedented insights into the atmospheres of planets across a wide range of masses and temperatures, revealing molecular compositions, thermal structures, and the prevalence of clouds and hazes \citep[][]{rust2023,powell2024,benneke2024,dy2024}. These results have firmly established atmospheric characterisation as a cornerstone for understanding planetary formation, evolution, and potential habitability.

This effort will be extended in the 2030s by ESA’s Ariel mission \citep{ariel2018}, conceived as the first large-scale atmospheric survey of exoplanets. By observing hundreds to thousands of planets, Ariel will provide a homogeneous baseline for comparative planetology and enable the identification of population-level trends. However, its moderate spectral resolution and wavelength coverage limit its ability to resolve key degeneracies, particularly for small planets where stellar contamination and overlapping molecular features dominate. As a result, Ariel’s contribution to the atmospheric characterisation of sub-Neptunes and terrestrial planets, especially on temperate orbits, will necessarily be restricted to a limited subset of favourable targets, preventing a comprehensive, population-level understanding of atmospheric diversity among small planets.

Extremely large ground-based telescopes will provide a complementary high-resolution view of exoplanet atmospheres. Facilities such as the ELT, with instruments like ANDES, will enable transformative studies of individual planets, extending atmospheric characterisation to smaller radii and, in favourable cases, to temperate rocky worlds \citep{palle2025}. High-resolution spectroscopy offers unique diagnostic power, including access to individual molecular lines and atmospheric dynamics, but at the cost of substantial observing time. Robust atmospheric detections for small planets typically require multiple transits or many hours of integration, restricting such observations to a small number of benchmark systems. Consequently, despite their extraordinary capabilities, ELT-class facilities are intrinsically unsuited for population-level atmospheric surveys. 

Without population-level atmospheric characterisation, particularly for sub-Neptunes and rocky planets, it will remain difficult to establish robust trends, characterise the diversity of atmospheric outcomes, and link them with planetary interiors. In this context, even the search for potential biosignatures will remain confined to a very limited number of favourable targets, preventing any statistically meaningful assessment of the prevalence of potentially inhabited worlds across planetary populations.

\subsection*{Fundamental Science Challenge}

The discussion presented above highlights a common and pervasive limitation across current and forthcoming exoplanet programmes. While the coming decades will deliver an unprecedented census of exoplanets across a wide range of radii, orbital periods, and stellar environments, there remains no clear pathway to translate these discoveries into a comprehensive, physically grounded understanding of planetary systems. Both planetary interiors and atmospheres, key tracers of formation pathways, evolutionary histories, and habitability, remain accessible only to a small, highly selected subset of planets.

The fundamental science challenge for the 2040s is therefore not one of discovery, but of interpretation at scale. Addressing how planets form, evolve, and acquire their present-day compositions requires the joint characterisation of planetary masses, interior structures, and atmospheric properties across statistically meaningful samples. Without such population-level physical characterisation, trends inferred from detections alone remain ambiguous and cannot be robustly linked to the underlying physical processes.

Meeting this challenge demands a shift in the observational paradigm, from detection-driven surveys and case-by-case studies to dedicated, large-scale efforts designed to deliver deep, homogeneous physical information. 

This motivates the need for a new class of facility with (i) greatly increased survey speed, 
(ii) scalable collecting area, and (iii) ultra-stable, cost-effective instrumentation capable of 
simultaneously supporting high-precision radial-velocity measurements and high-resolution 
atmospheric characterisation. In practice, this requires broad spectral coverage from the 
visible to the near-infrared ($\sim$0.5--1.8\,$\mu$m) at resolving powers of $R \gtrsim 100\,000$, 
combined with long-term instrumental stability at the cm\,s$^{-1}$ level.

\subsection*{Facility concept}

A promising route is a \textbf{photonics-enabled modular telescope architecture}, as exemplified by 
ongoing developments within the Photonic E-MARCOT consortium. Instead of a single monolithic 
aperture, a photonically combined fibre-fed modular array of telescopes, feeding photonic lanterns, pupil remapping devices, and compact photonic spectrographs, can deliver:
\begin{itemize}[leftmargin=0.5cm]
    \item An effective collecting area comparable to a $\sim 15$--30\,m telescope, with 
    significantly lower cost per square metre, enabling population-level surveys.
    \item Intrinsically stable, miniaturised, and replicable spectrographs providing 
    $R \gtrsim 100\,000$ over visible and near-infrared wavelengths, optimised for 
    cm\,s$^{-1}$ radial velocities and high-dispersion atmospheric spectroscopy.
    \item A scalable path to expand the array or replicate it at multiple sites, enabling 
    both increased survey speed and improved sky coverage.
\end{itemize}

In this way, the proposed science programme directly supports ESO's long-term vision of maintaining 
world-leading capabilities in ground-based astronomy well into the second half of the century.

%\newpage
\section*{References}
%\vspace{-1cm}
\renewcommand{\refname}{}

\setlength{\bibsep}{0pt} % no spacing between entries
\setlength{\bibhang}{0.5em} % smaller indentation

\begin{multicols}{2}
\begin{spacing}{0.9}
% \bibliographystyle{aa}
% \bibliography{ref}  % create refs.bib if needed

\begin{thebibliography}{15}
\expandafter\ifx\csname natexlab\endcsname\relax\def\natexlab#1{#1}\fi

\bibitem[{{Acu{\~n}a} {et~al.}(2024){Acu{\~n}a}, {Kreidberg}, {Zhai}, \& {Molli{\`e}re}}]{acuña2024}
{Acu{\~n}a}, L., {Kreidberg}, L., {Zhai}, M., \& {Molli{\`e}re}, P. 2024, Astronomy \& Astrophysics, 688, A60

\bibitem[{{Benneke} {et~al.}(2024){Benneke}, {Roy}, {Coulombe}, {Radica}, {Piaulet}, {Ahrer}, {Pierrehumbert}, {Krissansen-Totton}, {Schlichting}, {Hu}, {Yang}, {Christie}, {Thorngren}, {Young}, {Pelletier}, {Knutson}, {Miguel}, {Evans-Soma}, {Dorn}, {Gagnebin}, {Fortney}, {Komacek}, {MacDonald}, {Raul}, {Cloutier}, {Acuna}, {Lafreni{\`e}re}, {Cadieux}, {Doyon}, {Welbanks}, \& {Allart}}]{benneke2024}
{Benneke}, B., {Roy}, P.-A., {Coulombe}, L.-P., {et~al.} 2024, arXiv e-prints, arXiv:2403.03325

\bibitem[{{Borucki} {et~al.}(2010){Borucki}, {Koch}, {Basri}, {Batalha}, {Brown}, {Caldwell}, {Caldwell}, {Christensen-Dalsgaard}, {Cochran}, {DeVore}, {Dunham}, {Dupree}, {Gautier}, {Geary}, {Gilliland}, {Gould}, {Howell}, {Jenkins}, {Kondo}, {Latham}, {Marcy}, {Meibom}, {Kjeldsen}, {Lissauer}, {Monet}, {Morrison}, {Sasselov}, {Tarter}, {Boss}, {Brownlee}, {Owen}, {Buzasi}, {Charbonneau}, {Doyle}, {Fortney}, {Ford}, {Holman}, {Seager}, {Steffen}, {Welsh}, {Rowe}, {Anderson}, {Buchhave}, {Ciardi}, {Walkowicz}, {Sherry}, {Horch}, {Isaacson}, {Everett}, {Fischer}, {Torres}, {Johnson}, {Endl}, {MacQueen}, {Bryson}, {Dotson}, {Haas}, {Kolodziejczak}, {Van Cleve}, {Chandrasekaran}, {Twicken}, {Quintana}, {Clarke}, {Allen}, {Li}, {Wu}, {Tenenbaum}, {Verner}, {Bruhweiler}, {Barnes}, \& {Prsa}}]{kepler}
{Borucki}, W.~J., {Koch}, D., {Basri}, G., {et~al.} 2010, Science, 327, 977

\bibitem[{{Dorn} {et~al.}(2015){Dorn}, {Khan}, {Heng}, {Connolly}, {Alibert}, {Benz}, \& {Tackley}}]{dorn2015}
{Dorn}, C., {Khan}, A., {Heng}, K., {et~al.} 2015, Astronomy \& Astrophysics, 577, A83

\bibitem[{{Dyrek} {et~al.}(2024){Dyrek}, {Min}, {Decin}, {Bouwman}, {Crouzet}, {Molli{\`e}re}, {Lagage}, {Konings}, {Tremblin}, {G{\"u}del}, {Pye}, {Waters}, {Henning}, {Vandenbussche}, {Ardevol Martinez}, {Argyriou}, {Ducrot}, {Heinke}, {van Looveren}, {Absil}, {Barrado}, {Baudoz}, {Boccaletti}, {Cossou}, {Coulais}, {Edwards}, {Gastaud}, {Glasse}, {Glauser}, {Greene}, {Kendrew}, {Krause}, {Lahuis}, {Mueller}, {Olofsson}, {Patapis}, {Rouan}, {Royer}, {Scheithauer}, {Waldmann}, {Whiteford}, {Colina}, {van Dishoeck}, {{\"O}stlin}, {Ray}, \& {Wright}}]{dy2024}
{Dyrek}, A., {Min}, M., {Decin}, L., {et~al.} 2024, Nature, 625, 51

\bibitem[{{Ge} {et~al.}(2022){Ge}, {Zhang}, {Zang}, {Deng}, {Mao}, {Xie}, {Liu}, {Zhou}, {Willis}, {Huang}, {Howell}, {Feng}, {Zhu}, {Yao}, {Liu}, {Aizawa}, {Zhu}, {Li}, {Ma}, {Ye}, {Yu}, {Xiang}, {Yu}, {Liu}, {Yang}, {Wang}, {Shi}, {Fang}, {Zong}, {Liu}, {Zhang}, {Zhang}, {El-Badry}, {Shen}, {Tam}, {Hu}, {Yang}, {Zou}, {Wu}, {Lei}, {Wei}, {Wu}, {Sun}, {Wang}, {Zhang}, {Xu}, {Yang}, {Li}, {Xiang}, {Wang}, {Wang}, {Zhang}, {Jia}, {Yuan}, {Zhang}, {Xuesong Wang}, {Gan}, {Wang}, {Zhao}, {Liu}, {Wei}, {Kang}, {Yang}, {Qi}, {Liu}, {Zhang}, {Zhu}, {Zhou}, {Zhang}, {Yu}, {Zhang}, {Li}, {Tang}, {Wang}, {Wang}, {Li}, {Cheng}, {Shen}, {Li}, {Pan}, {Yang}, {Gao}, {Song}, {Wang}, {Zhang}, {Chen}, {Wang}, {Zhang}, {Wang}, {Zeng}, {Zheng}, {Zhu}, {Guo}, {Zhang}, {Li}, {Wen}, {Feng}, {Chen}, {Chen}, {Han}, {Yang}, {Wang}, {Duan}, {Huang}, {Liang}, {Bi}, {Gai}, {Ge}, {Guo}, {Huang}, {Li}, {Li}, {Li}, {Yuxi}, {Lu}, {Rix}, {Shi}, {Song}, {Tang}, {Ting}, {Wu}, {Wu}, {Yang}, {Yin}, {Gould}, {Lee}, {Dong}, {Yee}, {Shvartzvald},
  {Yang}, {Kuang}, {Zhang}, {Liao}, {Qi}, {Yang}, {Zhang}, {Jiang}, {Ou}, {Li}, {Beck}, {Bedding}, {Campante}, {Chaplin}, {Christensen-Dalsgaard}, {Garc{\'\i}a}, {Gaulme}, {Gizon}, {Hekker}, {Huber}, {Khanna}, {Li}, {Mathur}, {Miglio}, {Mosser}, {Ong}, {Santos}, {Stello}, {Bowman}, {Lares-Martiz}, {Murphy}, {Niu}, {Ma}, {Moln{\'a}r}, {Fu}, {De Cat}, {Su}, \& {consortium}}]{earth2022}
{Ge}, J., {Zhang}, H., {Zang}, W., {et~al.} 2022, arXiv e-prints, arXiv:2206.06693

\bibitem[{{Howard}(2013)}]{howard2013}
{Howard}, A.~W. 2013, Science, 340, 572

\bibitem[{{Palle} {et~al.}(2025){Palle}, {Biazzo}, {Bolmont}, {Molli{\`e}re}, {Poppenhaeger}, {Birkby}, {Brogi}, {Chauvin}, {Chiavassa}, {Hoeijmakers}, {Lellouch}, {Lovis}, {Maiolino}, {Nortmann}, {Parviainen}, {Pino}, {Turbet}, {Weder}, {Albrecht}, {Antoniucci}, {Barros}, {Beaudoin}, {Benneke}, {Boisse}, {Bonomo}, {Borsa}, {Brandeker}, {Brandner}, {Buchhave}, {Cheffot}, {Deborde}, {Debras}, {Doyon}, {Di Marcantonio}, {Giacobbe}, {Gonz{\'a}lez Hern{\'a}ndez}, {Helled}, {Kreidberg}, {Machado}, {Maldonado}, {Marconi}, {Martins}, {Miceli}, {Mordasini}, {N'Diaye}, {Niedzielski}, {Nisini}, {Origlia}, {Peroux}, {Pietrow}, {Pinna}, {Rauscher}, {Reffert}, {Rodr{\'\i}guez-L{\'o}pez}, {Rousselot}, {Sanna}, {Santos}, {Simonnin}, {Su{\'a}rez Mascare{\~n}o}, {Zanutta}, {Zapatero-Osorio}, \& {Zechmeister}}]{palle2025}
{Palle}, E., {Biazzo}, K., {Bolmont}, E., {et~al.} 2025, Experimental Astronomy, 59, 29

\bibitem[{{Plotnykov} \& {Valencia}(2024)}]{plot2024}
{Plotnykov}, M. \& {Valencia}, D. 2024, Monthly Notices of the Royal Astronomical Society, 530, 3488

\bibitem[{{Powell} {et~al.}(2024){Powell}, {Feinstein}, {Lee}, {Zhang}, {Tsai}, {Taylor}, {Kirk}, {Bell}, {Barstow}, {Gao}, {Bean}, {Blecic}, {Chubb}, {Crossfield}, {Jordan}, {Kitzmann}, {Moran}, {Morello}, {Moses}, {Welbanks}, {Yang}, {Zhang}, {Ahrer}, {Bello-Arufe}, {Brande}, {Casewell}, {Crouzet}, {Cubillos}, {Demory}, {Dyrek}, {Flagg}, {Hu}, {Inglis}, {Jones}, {Kreidberg}, {L{\'o}pez-Morales}, {Lagage}, {Meier Vald{\'e}s}, {Miguel}, {Parmentier}, {Piette}, {Rackham}, {Radica}, {Redfield}, {Stevenson}, {Wakeford}, {Aggarwal}, {Alam}, {Batalha}, {Batalha}, {Benneke}, {Berta-Thompson}, {Brady}, {Caceres}, {Carter}, {D{\'e}sert}, {Harrington}, {Iro}, {Line}, {Lothringer}, {MacDonald}, {Mancini}, {Molaverdikhani}, {Mukherjee}, {Nixon}, {Oza}, {Palle}, {Rustamkulov}, {Sing}, {Steinrueck}, {Venot}, {Wheatley}, \& {Yurchenko}}]{powell2024}
{Powell}, D., {Feinstein}, A.~D., {Lee}, E. K.~H., {et~al.} 2024, Nature, 626, 979

\bibitem[{{Rauer} {et~al.}(2025){Rauer}, {Aerts}, {Cabrera}, {Deleuil}, {Erikson}, {Gizon}, {Goupil}, {Heras}, {Walloschek}, {Lorenzo-Alvarez}, {Marliani}, {Martin-Garcia}, {Mas-Hesse}, {O'Rourke}, {Osborn}, {Pagano}, {Piotto}, {Pollacco}, {Ragazzoni}, {Ramsay}, {Udry}, {Appourchaux}, {Benz}, {Brandeker}, {G{\"u}del}, {Janot-Pacheco}, {Kabath}, {Kjeldsen}, {Min}, {Santos}, {Smith}, {Suarez}, {Werner}, {Aboudan}, {Abreu}, {Acu{\~n}a}, {Adams}, {Adibekyan}, {Affer}, {Agneray}, {Agnor}, {Aguirre B{\o}rsen-Koch}, {Ahmed}, {Aigrain}, {Al-Bahlawan}, {Alcacera Gil}, {Alei}, {Alencar}, {Alexander}, {Alfonso-Garz{\'o}n}, {Alibert}, {Allende Prieto}, {Almeida}, {Alonso Sobrino}, {Altavilla}, {Althaus}, {Alvarez Trujillo}, {Amarsi}, {Ammler-von Eiff}, {Am{\^o}res}, {Andrade}, {Antoniadis-Karnavas}, {Ant{\'o}nio}, {Aparicio del Moral}, {Appolloni}, {Arena}, {Armstrong}, {Aroca Aliaga}, {Asplund}, {Audenaert}, {Auricchio}, {Avelino}, {Baeke}, {Bailli{\'e}}, {Balado}, {Ballber Balaguer{\'o}}, {Balestra}, {Ball}, {Ballans},
  {Ballot}, {Barban}, {Barbary}, {Barbieri}, {Barcel{\'o} Forteza}, {Barker}, {Barklem}, {Barnes}, {Barrado Navascues}, {Barragan}, {Baruteau}, {Basu}, {Baudin}, {Baumeister}, {Bayliss}, {Bazot}, {Beck}, {Belkacem}, {Bellinger}, {Benatti}, {Benomar}, {B{\'e}rard}, {Bergemann}, {Bergomi}, {Bernardo}, {Biazzo}, {Bignamini}, {Bigot}, {Billot}, {Binet}, {Biondi}, {Biondi}, {Birch}, {Bitsch}, {Bluhm Ceballos}, {B{\'o}di}, {Bogn{\'a}r}, {Boisse}, {Bolmont}, {Bonanno}, {Bonavita}, {Bonfanti}, {Bonfils}, {Bonito}, {Bonomo}, {B{\"o}rner}, {Boro Saikia}, {Borreguero Mart{\'\i}n}, {Borsa}, {Borsato}, {Bossini}, {Bouchy}, {Bou{\'e}}, {Boufleur}, {Boumier}, {Bourrier}, {Bowman}, {Bozzo}, {Bradley}, {Bray}, {Bressan}, {Breton}, {Brienza}, {Brito}, {Brogi}, {Brown}, {Brown}, {Brun}, {Bruno}, {Bruns}, {Buchhave}, {Bugnet}, {Buldgen}, {Burgess}, {Busatta}, {Busso}, {Buzasi}, {Caballero}, {Cabral}, {Cabrero Gomez}, {Calderone}, {Cameron}, {Cameron}, {Campante}, {Campos Gestal}, {Canto Martins}, {Cara}, {Carone}, {Carrasco},
  {Casagrande}, {Casewell}, {Cassisi}, {Castellani}, {Castro}, {Catala}, {Catal{\'a}n Fern{\'a}ndez}, {Catelan}, {Cegla}, {Cerruti}, {Cessa}, {Chadid}, {Chaplin}, {Charpinet}, {Chiappini}, {Chiarucci}, {Chiavassa}, {Chinellato}, {Chirulli}, {Christensen-Dalsgaard}, {Church}, {Claret}, {Clarke}, {Claudi}, {Clermont}, {Coelho}, {Coelho}, {Cogato}, {Colom{\'e}}, {Condamin}, {Conde Garc{\'\i}a}, \& {Conseil}}]{plato2025}
{Rauer}, H., {Aerts}, C., {Cabrera}, J., {et~al.} 2025, Experimental Astronomy, 59, 26

\bibitem[{{Ricker} {et~al.}(2015){Ricker}, {Winn}, {Vanderspek}, {Latham}, {Bakos}, {Bean}, {Berta-Thompson}, {Brown}, {Buchhave}, {Butler}, {Butler}, {Chaplin}, {Charbonneau}, {Christensen-Dalsgaard}, {Clampin}, {Deming}, {Doty}, {De Lee}, {Dressing}, {Dunham}, {Endl}, {Fressin}, {Ge}, {Henning}, {Holman}, {Howard}, {Ida}, {Jenkins}, {Jernigan}, {Johnson}, {Kaltenegger}, {Kawai}, {Kjeldsen}, {Laughlin}, {Levine}, {Lin}, {Lissauer}, {MacQueen}, {Marcy}, {McCullough}, {Morton}, {Narita}, {Paegert}, {Palle}, {Pepe}, {Pepper}, {Quirrenbach}, {Rinehart}, {Sasselov}, {Sato}, {Seager}, {Sozzetti}, {Stassun}, {Sullivan}, {Szentgyorgyi}, {Torres}, {Udry}, \& {Villasenor}}]{ricker2015}
{Ricker}, G.~R., {Winn}, J.~N., {Vanderspek}, R., {et~al.} 2015, Journal of Astronomical Telescopes, Instruments, and Systems, 1, 014003

\bibitem[{{Rustamkulov} {et~al.}(2023){Rustamkulov}, {Sing}, {Mukherjee}, {May}, {Kirk}, {Schlawin}, {Line}, {Piaulet}, {Carter}, {Batalha}, {Goyal}, {L{\'o}pez-Morales}, {Lothringer}, {MacDonald}, {Moran}, {Stevenson}, {Wakeford}, {Espinoza}, {Bean}, {Batalha}, {Benneke}, {Berta-Thompson}, {Crossfield}, {Gao}, {Kreidberg}, {Powell}, {Cubillos}, {Gibson}, {Leconte}, {Molaverdikhani}, {Nikolov}, {Parmentier}, {Roy}, {Taylor}, {Turner}, {Wheatley}, {Aggarwal}, {Ahrer}, {Alam}, {Alderson}, {Allen}, {Banerjee}, {Barat}, {Barrado}, {Barstow}, {Bell}, {Blecic}, {Brande}, {Casewell}, {Changeat}, {Chubb}, {Crouzet}, {Daylan}, {Decin}, {D{\'e}sert}, {Mikal-Evans}, {Feinstein}, {Flagg}, {Fortney}, {Harrington}, {Heng}, {Hong}, {Hu}, {Iro}, {Kataria}, {Kempton}, {Krick}, {Lendl}, {Lillo-Box}, {Louca}, {Lustig-Yaeger}, {Mancini}, {Mansfield}, {Mayne}, {Miguel}, {Morello}, {Ohno}, {Palle}, {Petit dit de la Roche}, {Rackham}, {Radica}, {Ramos-Rosado}, {Redfield}, {Rogers}, {Shkolnik}, {Southworth}, {Teske}, {Tremblin},
  {Tucker}, {Venot}, {Waalkes}, {Welbanks}, {Zhang}, \& {Zieba}}]{rust2023}
{Rustamkulov}, Z., {Sing}, D.~K., {Mukherjee}, S., {et~al.} 2023, Nature, 614, 659

\bibitem[{{Tinetti} {et~al.}(2018){Tinetti}, {Drossart}, {Eccleston}, {Hartogh}, {Heske}, {Leconte}, {Micela}, {Ollivier}, {Pilbratt}, {Puig}, {Turrini}, {Vandenbussche}, {Wolkenberg}, {Beaulieu}, {Buchave}, {Ferus}, {Griffin}, {Guedel}, {Justtanont}, {Lagage}, {Machado}, {Malaguti}, {Min}, {N{\o}rgaard-Nielsen}, {Rataj}, {Ray}, {Ribas}, {Swain}, {Szabo}, {Werner}, {Barstow}, {Burleigh}, {Cho}, {Coud{\'e} du Foresto}, {Coustenis}, {Decin}, {Encrenaz}, {Galand}, {Gillon}, {Helled}, {Morales}, {Garc{\'\i}a Mu{\~n}oz}, {Moneti}, {Pagano}, {Pascale}, {Piccioni}, {Pinfield}, {Sarkar}, {Selsis}, {Tennyson}, {Triaud}, {Venot}, {Waldmann}, {Waltham}, {Wright}, {Amiaux}, {Augu{\`e}res}, {Berth{\'e}}, {Bezawada}, {Bishop}, {Bowles}, {Coffey}, {Colom{\'e}}, {Crook}, {Crouzet}, {Da Peppo}, {Sanz}, {Focardi}, {Frericks}, {Hunt}, {Kohley}, {Middleton}, {Morgante}, {Ottensamer}, {Pace}, {Pearson}, {Stamper}, {Symonds}, {Rengel}, {Renotte}, {Ade}, {Affer}, {Alard}, {Allard}, {Altieri}, {Andr{\'e}}, {Arena}, {Argyriou},
  {Aylward}, {Baccani}, {Bakos}, {Banaszkiewicz}, {Barlow}, {Batista}, {Bellucci}, {Benatti}, {Bernardi}, {B{\'e}zard}, {Blecka}, {Bolmont}, {Bonfond}, {Bonito}, {Bonomo}, {Brucato}, {Brun}, {Bryson}, {Bujwan}, {Casewell}, {Charnay}, {Pestellini}, {Chen}, {Ciaravella}, {Claudi}, {Cl{\'e}dassou}, {Damasso}, {Damiano}, {Danielski}, {Deroo}, {Di Giorgio}, {Dominik}, {Doublier}, {Doyle}, {Doyon}, {Drummond}, {Duong}, {Eales}, {Edwards}, {Farina}, {Flaccomio}, {Fletcher}, {Forget}, {Fossey}, {Fr{\"a}nz}, {Fujii}, {Garc{\'\i}a-Piquer}, {Gear}, {Geoffray}, {G{\'e}rard}, {Gesa}, {Gomez}, {Graczyk}, {Griffith}, {Grodent}, {Guarcello}, {Gustin}, {Hamano}, {Hargrave}, {Hello}, {Heng}, {Herrero}, {Hornstrup}, {Hubert}, {Ida}, {Ikoma}, {Iro}, {Irwin}, {Jarchow}, {Jaubert}, {Jones}, {Julien}, {Kameda}, {Kerschbaum}, {Kervella}, {Koskinen}, {Krijger}, {Krupp}, {Lafarga}, {Landini}, {Lellouch}, {Leto}, {Luntzer}, {Rank-L{\"u}ftinger}, {Maggio}, {Maldonado}, {Maillard}, {Mall}, {Marquette}, {Mathis}, {Maxted}, {Matsuo},
  {Medvedev}, {Miguel}, {Minier}, {Morello}, {Mura}, {Narita}, {Nascimbeni}, {Nguyen Tong}, {Noce}, {Oliva}, {Palle}, {Palmer}, {Pancrazzi}, {Papageorgiou}, {Parmentier}, {Perger}, {Petralia}, {Pezzuto}, {Pierrehumbert}, \& {Pillitteri}}]{ariel2018}
{Tinetti}, G., {Drossart}, P., {Eccleston}, P., {et~al.} 2018, Experimental Astronomy, 46, 135

\bibitem[{{Winn} \& {Fabrycky}(2015)}]{winn2015}
{Winn}, J.~N. \& {Fabrycky}, D.~C. 2015, Annual Review of Astronomy and Astrophysics, 53, 409

\end{thebibliography}

\end{spacing}
\end{multicols}

\end{document}